# Usability, Design and Content Issues of Mobile Apps for Cultural Heritage Promotion: The Malta Culture Guide Experience

Stefania Boiano
InvisibleStudio
Unit 36, 88–90 Hatton Garden
London EC1N 8PN
United Kingdom
marketing@invisblestudio.it
www.stefaniaboiano.net

Jonathan Bowen
London South Bank University
Faculty of Business, Borough Road
London SE1 0AA
United Kingdom
jonathan.bowen@lsbu.ac.uk
www.jpbowen.com

Giuliano Gaia
InvisibleStudio
Via Circo 1
20123 Milan
Italy
marketing@invisblestudio.it
www.giulianogaia.net

**The paper discusses the experience of producing and distributing an iPhone app for promotion of the Maltese Cultural Heritage on behalf of the Malta Tourism Authority. Thanks to its position at the heart of the Mediterranean Sea, Malta has been a crossroads of civilisations whose traces are still visible today, leaving a particularly rich and varied cultural heritage, from megalithic temples to baroque palaces and Caravaggio masterpieces. Conveying all these different aspects within a single application, using textual, visual, and audio means, has raised many different issues about the planning and production of cultural content for mobile usage, together with usability aspects regarding design and distribution of a mobile app. In this paper, we outline all of these aspects, focusing on the design and planning strategies for a long-term user commitment and how to evaluate results for cultural mobile applications. We include experience of all the steps of developing a mobile app, information that is of possible benefit to other app developers in the cultural sector.**

*Mobile apps. Usability. App design. Multimedia. Cultural heritage.*

.

## 1. INTRODUCTION

The use of mobile devices is increasing dramatically in the cultural and museum sectors (Filippini-Fantoni & Bowen 2008). Information on this area of application is available online (e.g., http://museummobile.info).

One approach is to provide customised devices to visitors. An example is the visitor guide available at the British Museum (Filippini-Fantoni et al. 2011, McDaid et al. 2011). This can have facilities in several languages, helping to avoid multi-language museum labels, for example.

Another approach is to use the smartphones of the visitors themselves, by providing a suitable mobile app that can be downloaded onto the device. This could be for an open system such as Android or a closed system with associated restrictions such as the Apple iPhone/iPad (Keene 2011).

In this paper, we will consider a particular cultural example of a mobile app, providing information that should be of help to developers of similar apps.

## 2. PLANNING A MOBILE APP

Before planning a mobile app, one should take into consideration different aspects, such as:

- Should it be a mobile app or is a web-based site aimed at mobile use good enough?
- For which platforms should the app be developed?
- Is the content "mobile ready"?
- How are maintenance and updates planned to be done?

We will now briefly discuss all of these aspects.

## 2.1 Mobile apps vs. mobile sites

This is a basic question that every institution willing to develop a mobile app should consider (Forbes 2011). Mobile websites, i.e., websites optimised to be used by mobile devices, offer many advantages, such as:

- Optimising a website is usually cheaper than developing a new app.
- Platform-independence: one website is suitable for all devices (iPhone, iPad, Android, BlackBerry, etc.).
- Only the website content has to be updated.
- The technology required is easier and mainstream, such as PHP, (X)HTML, and CSS.
- There is independence from app stores publication policies, since it is not necessary to publish an app in a store like the Apple app store. Instead, the website can be published and nothing else is required. Publication on the Apple app store can be a long and painful process (Keene 2011).

On the other hand, apps at the moment offer better usability for the end user than mobile websites, as stated by the well-known usability researcher Jakob Nielsen:

*"As of this writing, there's no contest: ship **mobile apps** if you can afford it. Our usability studies with mobile devices clearly show that users perform better with apps than with mobile sites. (Mobile sites have higher measured usability than desktop/full sites when used on a phone, but mobile apps score even higher.)"* (Nielsen 2012).

This happens because, according to Nielsen (and from our experience):

*"An **app can target the specific limitations and abilities of each individual device** much better than a website can while running inside a browser."* (Nielsen 2012).

For example, an app can make better use of device-specific features, such as GPS positioning, compass and camera. Overall mobile browsers offer an impoverished user experience while navigating websites, even mobile-optimised ones.

## 2.2 Choosing the platform

When developing a device-specific app, it is necessary to decide on all the different devices that are going to be supported. The first thing to take into consideration is market share. According to the NielsenWire blog by Nielsen Media Rating (NielsenWire 2012), the US mobile market share at the end of 2011 was as shown in Table 1 below.

*Table 1:* US OS Mobile Market Share (February 2012).

| OS | US mobile market share | |
|---|---|---|
| | All smartphone owners | Recent smartphone acquirers (3 months) |
| Android | 48% | 48% |
| iOS (iPhone, iPad, iPod Touch) | 32% | 43% |
| RIM BlackBerry | 12% | 5% |
| Other | 8% | 4% |

Younger people tend to have smartphones, whereas older people tend to have more tradition mobile technology (Beasley & Conway 2012). Although figures may vary from country to country, one thing which is quite consistent among the various markets is that the main smartphone platforms are Android and iOS (Apple's mobile operating system).

When we started developing the Malta Culture Guide app in 2010, the iPhone held an overwhelming majority of the smartphone market share, so it was an easy choice. As Table 1 shows, the situation is quite different now. Thus, we recommend whenever possible to develop for both Android and iOS, avoiding support for only one platform with apps intended for general use. On the other hand, it is now easier to ignore other platforms, unless of course one is targeting a niche market (for example if developing a corporate app for employees who often use BlackBerry smartphones).

## 2.3 Selecting the content

As obvious as it may sound, it should always be remembered that content is the single most important feature of every cultural app. This means content has to be engaging, worthwhile, and in particular suitable for mobile use.

Consider some basic questions about the content. Since many apps are derived from pre-existing content, one should start by analysing this content and asking some basic questions, such as:

i. Is it valuable content?
ii. Is it easy to read/watch/listen to and understand the content? When using a mobile device, users are often in a noisy and distracting environment, so information should be even easier than usual web content aimed at desktop use.

iii. Does it require updates? Updating can be a tricky aspect of mobile apps (see the relevant section later in this paper).
iv. Does it make use of the device features? For example, any information that can be shown on a Google Map is usually very useful, because maps and "near me" navigation are helpful features that make an app more appreciated by users.

## 3. CONTENT TYPOLOGIES

We will now briefly consider different types of content relating to their mobile use.

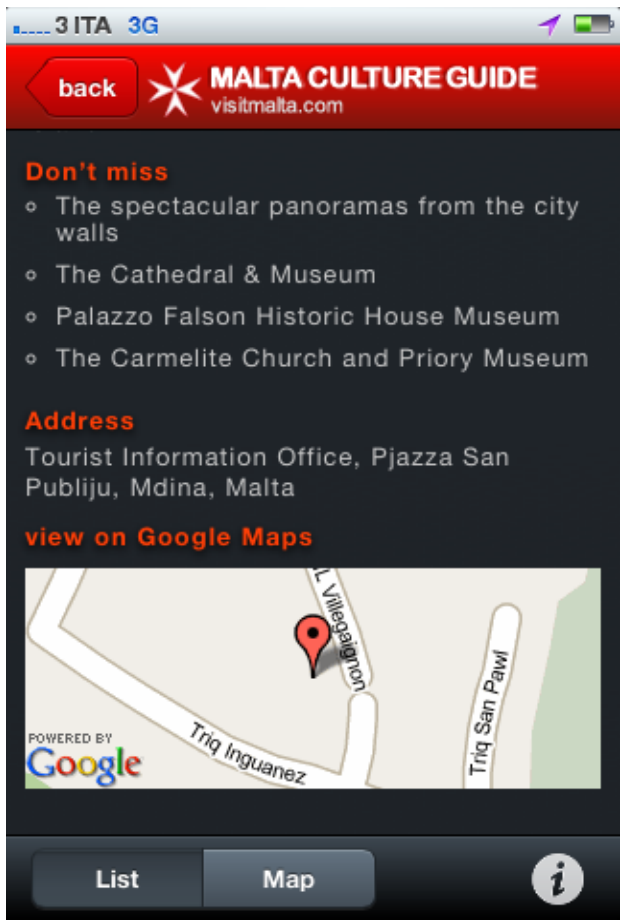

*Figure 1: Example of text from the Malta Culture Guide.*

### 3.1 Text

When considering text for mobile usage, the golden rule is to keep it as short as possible. Web usability studies (Nielsen 1997) show that people on PCs usually do not read, but scan. This is even truer on mobile devices with small screens and poor contrast, often used outdoors. As Nielsen findings suggest:

*"User comprehension scores on the Cloze test were 48% of the desktop level when using the iPhone-sized screen. That is, it's roughly twice as hard to understand complicated content when reading on the smaller screen."* (Nielsen 2012).

This means, as Nielsen himself concludes, that for the user "Mobile content is twice as difficult". Therefore it is advisable to break text into small paragraphs, keep sentences short, and use subtitles and lists to make scanning easier. Longer text may be considered for applications aimed at tablets (like the Apple iPad), which have bigger screens and are more for "sofa" than "walking" use.

### 3.2 Images

Images suffer the same problem as text when used on mobile devices, due to the limited size and quality of the screen. This means that users will not be able to appreciate details (for example of a painting or monument). If it is necessary to show a detail, it can be useful to enlarge it as a separate image.

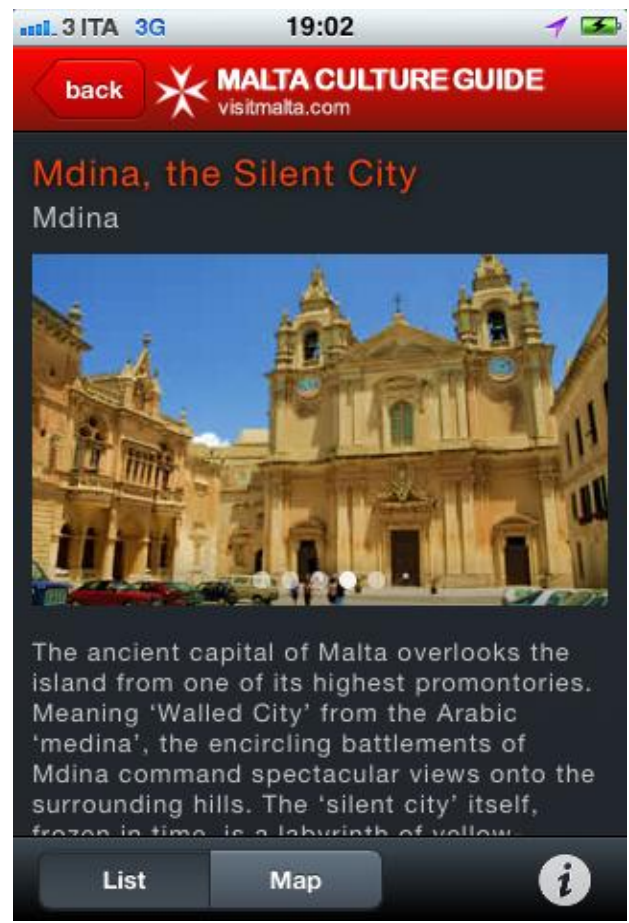

*Figure 2: Example of detail image from the Malta Culture Guide.*

High-contrast images are better and it is preferable to focus on a few elements rather than use an image full of different details. Again, tablets offer a much better visual experience and one could consider using higher resolution images for a tablet version of an app.

### 3.3 Audio

Audio is excellent content for mobile devices, because it is ideal for mobile use and earphones can offer a satisfying user experience. Mobile apps can be a very good way of reusing valuable audio content that is already available. As Nancy Proctor writes:

> *"Other than audio tours loaned out on made-for-museums devices, podcasts are probably the most common mobile media being published by Museums."* (Proctor 2011).

The opportunity to embed this existing content into a new environment for the user to enjoy should not be missed by cultural institutions. For example, we have included in the Malta Culture Guide app some podcasts (Boiano & Gaia 2006) about Maltese UNESCO World Heritage sites that we had produced the previous year.

When producing or considering audio content, bear in mind that audio files should be not too long and they should be clearly audible without too much background music, because outdoor use can already provide a noisy background, even if the user is wearing earphones.

One should also carefully take into account the file size, because one should aim at making the app as small as possible to help speed up its download. For example, Apple has a strict policy that forces any application bigger than 20 Megabytes to be downloaded only if there is a wireless connection, making it impossible to download it on a 3G network connection. This threshold should be kept in mind when thinking about the content, because audio files can quickly exceed this size limit.

### 3.4 Video

Video suffers of the same limitations as images: restricted screen size and quality impoverish the user experience. Nevertheless, short videos are normally appreciated by users and can add an emotional layer to the cultural information. Videos distract the user from the surrounding environment, so they should be used carefully when the app is planned for usage in the galleries or outdoor, while they are usually much more suitable for use on tablets with their larger screen size.

Figure 3 shows an example of video content on the app.

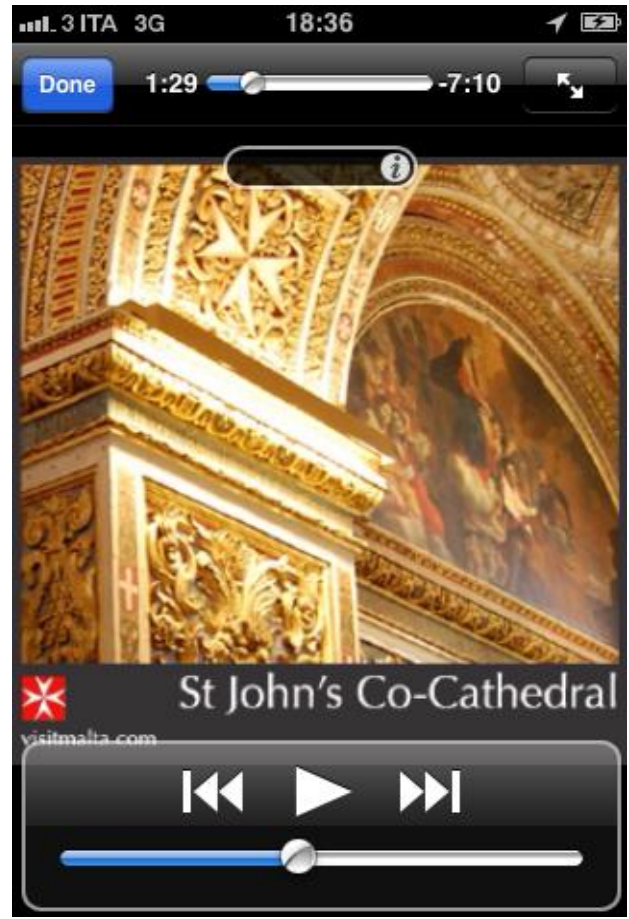

***Figure 3:*** *Example of audio/video content from the Malta Culture Guide.*

### 3.5 Maps

Interactive maps are one of the most successful features of smartphones, being very useful for showing location-based information. Google Maps (http://maps.google.com**)** provides an excellent tool, but one should consider carefully whether to include an offline map as well, in order to avoid roaming costs.

In the case of the Malta Culture Guide, our main target was foreign tourists visiting the islands, so we provided them with an offline map that was based on the free wiki world map OpenStreetMap (http://www.openstreetmap.org**)**, a leading open-source map. We also added links to the standard Google Map for users with suitable data plans or connected to a WiFi network (for example, when in a hotel).

Figure 4 on the next page shows an example of a map on the app.

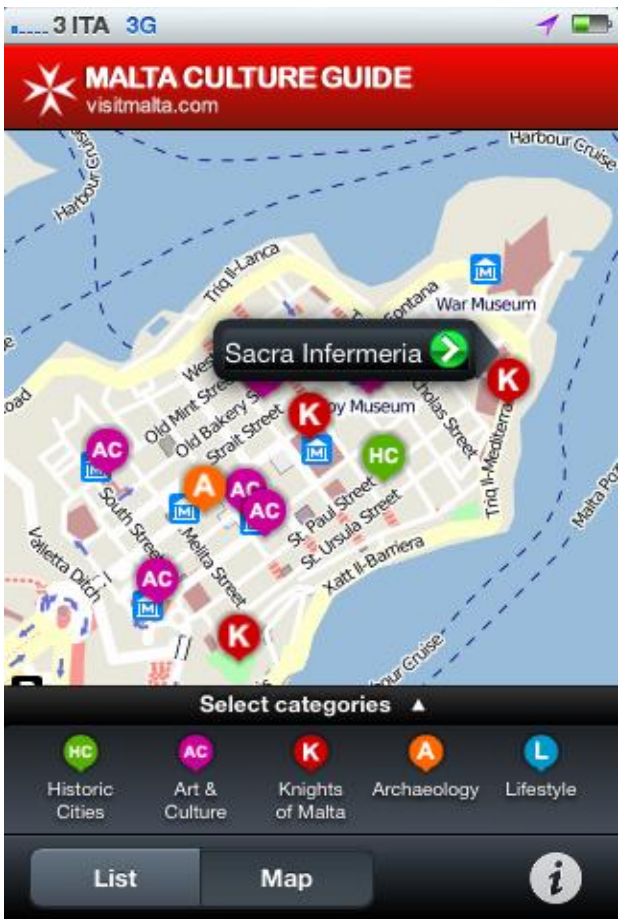

*Figure 4: The offline map from the Malta Culture Guide.*

### 3.6 "Social" features

According to Facebook statistics, nearly half of its 800 million users in 2011 used the social network via a mobile device. Apple has recently inserted Twitter as a native app into its iPhone and iPad. More and more apps take advantage of these features, enabling the user to share specific content via Facebook, email, Twitter, or other social networks. One should therefore plan in advance an easy way for users to share content, in order to facilitate diffusion and promotion of the app.

### 3.7 Updates and maintenance

The updating process should be carefully planned in advance, since every time the content is changed, users have to download the whole app from the app store again, an operation that can be annoying or forgotten. It is therefore advisable to create an automatic updating system that regularly checks and downloads content updates from the web without forcing the user to download and reinstall the whole app. One should also be aware that updates to the device operating system and the release of new models can create problems, so one should plan for a scheduled testing of the app functionalities at least every six months (or at every major OS update).

## 4. MOBILE DESIGN

When designing a mobile app the *user experience* design (UX design) is particularly important, because given the limitations of the device, usability is crucial for any app success.

First of all we produced a series of quick pen sketches of the main screens in order to understand the main elements to be taken into consideration. For example, see Figure 5 below.

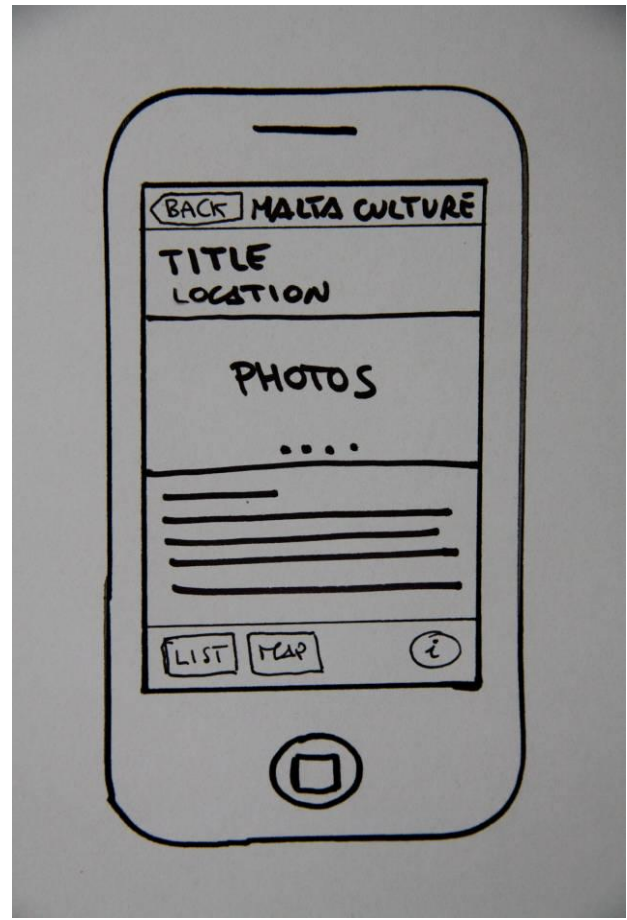

*Figure 5: Production sketch for the Malta Culture Guide.*

Then we produced some paper mockups. Mockups are an excellent way to work on the mobile user experience. In the case of the Malta Culture Guide we used paper mockups with sticky "post-it" notes on them, in order to design the full navigation before starting to work on the graphic layouts.

Coloured sticky notes are especially useful for outlining the information architecture in an easy and collaborative way. For some examples of these paper mockups, see Figure 6 below.

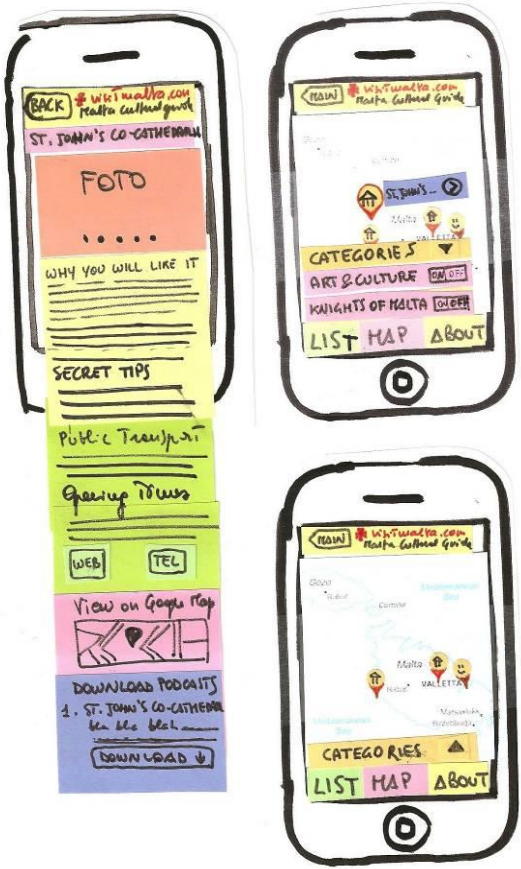

*Figure 6: Paper mock-ups
of the Malta Culture Guide.*

After the mock-ups with sticky post-it notes, we started working on the graphic digital layouts, always with a keen eye on text readability, including particularly in outdoor conditions. Menus should be made as short as possible and one should take advantage of the standard conventions such as image carousels and the provision of a "Back" button.

As a general rule, we suggest keeping the user interface as simple and intuitive, as possible. It is important to be aware of the fact that fingers are an imprecise pointing device. It is therefore advisable to keep the clickable area around links large enough to be selected easily. This helps to make the app as accessible as possible, especially for disabled or older people who may have difficulty using technology (Bowen 2005).

For a screenshot of the navigation menu, see Figure 7.

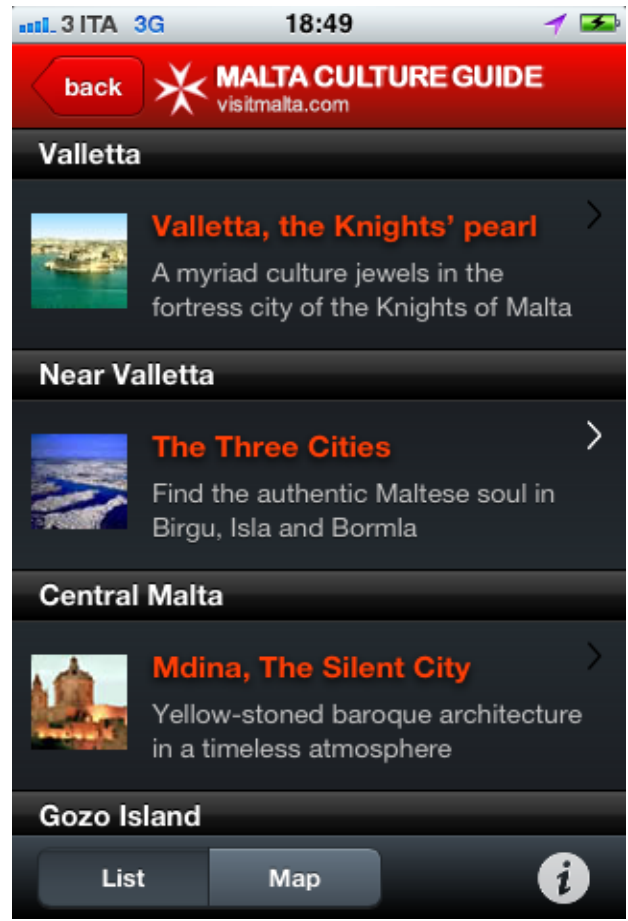

*Figure 7: Navigation menu
from the Malta Culture Guide.*

## 5. DEVELOPMENT AND MONITORING

In the development stage, it should first be decided whether to develop the app in the device's native language, such as Xcode for iPhone/iPad or Java for Android, or to use a framework. Using a framework, like Phonegap (http://phonegap.com), it is possible to develop in HTML 5 and JavaScript and let the software compile the app for the developer. Frameworks have pros and cons. On the plus side, they can significantly speed up the process of creating standard applications. On the negative side, they lack the flexibility of developing in native code and are subject to continuous updates. This means that it is necessary to check carefully if all the features are still functioning every time the app is modified.

As for monitoring of app use, app stores offer only limited data, such as the quantity of downloads and the country where the app has been downloaded. Luckily, more sophisticated analytics solutions are becoming available. For example Google Analytics (http://www.google.com/analytics), a popular web analytics tool, has developed a mobile version capable of tracking user behaviours when using an iOS or Android app.

Finally, user reviews on the app store should be checked carefully, because they provide important qualitative data and can orientate other users' choices.

## 6. CONCLUSIONS

This paper has provided a guide on how to develop a mobile app in the cultural sector, including practical information for others who may wish to do so. A case study of a real guide for tourists (in Malta) for the iPhone has been presented. Consideration of the choice between an app and a mobile website as well as the choice of platform and content has been included. Design, development, and monitoring aspects have also been covered.

For the future, more personalisation features could be added to such apps, as is available on advanced websites (Bowen & Filippini-Fantoni 2004). The HCI design could be adapted for different user groups, with accessibility in mind (Boiano et al. 2008, Bowen 2005).

The web is rapidly adapting to provide an increasing sense of community, including in cultural and museum spheres (Beler et al. 2004). Social networking and collaborative features could be integrated within an app, providing users with the possibility of collaborative feedback and interaction (Borda & Bowen 2011, Liu & Bowen 2011). For example, blog-style interaction currently available on the web could be adapted in the form of an app for existing culturally oriented blogs and online magazines (Beazley et al. 2010, Liu et al. 2010).

In summary, it is believed that there will be an increasing number of cultural apps and that these will become more interactive in nature, allowing user generated content as well as professionally written content.

**Acknowledgements**

*The Malta Culture Guide is downloadable from:*

*www.maltacultureguide.com*

*Special thanks to Dominic Micallef and Claude Zammit-Trevisan of the Malta Tourism Authority and Massimiliano Canestrari, who joined InvisibleStudio (http://www.invisiblestudio.it) team for the app development.*

*Jonathan Bowen thanks Museophile Limited (http://www.museophile.com) for support.*